\documentstyle[12pt]{article}

\newcommand{\nS}{{\bf S}}
\newcommand{\h}{\frac{1}{2}}

\title
{\bf\large\bf A NEW VISTA TO THE $SU(N)$ GAUGE FIELD
CONFINEMENT$^{\star}$}
\author{{\bf\normalsize\bf Zheng-Kun Zhu}\\
{\small\em Department of Physics and Atmospheric Science, Drexel University}\\
      {\small\em Philadelphia, Pennsyvalnia 19104-9984}\\
{\small and}\\
{\bf\normalsize\bf Da Hsuan Feng}\\
{\small\em Department of Physics and Atmospheric Science,
Drexel University}\\
      {\small\em Philadelphia, Pennsyvalnia 19104-9984}\\
{\small\em and Physics Division,
Oak Ridge National Laboratory, Oak Ridge, Tennessee 37996}
}
\date{\normalsize  }

\begin{document}

\maketitle

\begin{abstract}
Based on our exact effective spin model, the zero temperature $SU(N)$ gauge
field confinement has been studied by the Wilson line, rather than the quark
antiquark potential. The procedure may allow the confinement be studied in the
true continuum limit. For the $(1+1)$ $SU(2)$ gauge field, we have shown that
the confinement (here we mean that the Wilson line vanishes.) corresponds to
a linear zero temperature Polyakov quark antiquark potential. Also,
for the first time, the Polyakov and Wilson potentials are shown to be
identical.
\begin{flushleft}
\end{flushleft}
\end{abstract}

\newpage

A substantial amount of numerical evidences have been accumulated in the past
decade which indicate that the zero temperature $(3+1)$ dimensional $SU(2)$ and
$SU(3)$ gauge fields could indeed be permanently confined \cite{creutz}. Still,
one may argue that such a proof is not quite within the realm of the true
continuum world \cite{seiler}, and therefore a direct study of the confinement
in the continuum world is still highly desirable. In this paper, we shall
introduce a new approach which appears to have this feasibility.

Let us first consider the confinement of the finite temperature $SU(N)$ gauge
field.  It was speculated as early as the late seventies that in this case
there is a transition from the color confined to the color deconfined phase
with the Wilson line as its order parameter. Although this suggests that it is
natural to use the Wilson line to study the zero temperature
$SU(N)$ confinement by checking whether it is
zero or not, it was not carried out as far as we are aware of
\cite{ex1}.  Recently, the exact form of the effective spin model of the
$SU(N)$ gauge field was derived \cite{zhu}. For the zero temperature lattice
gauge theory (LGT), we found that
one can in principle compute the coupling constant of the effective spin model
because the effective action for the space-like link can be obtained exactly. Hence
by computing the coupling constant of the effective spin model in the continuum
limit, one can determine whether or not the gauge field confines since the $3$
dimensional effective spin model can be easily carried out numerically. This
thus opens up a possibility to study the zero temperature confinement by the
Wilson line.  In this paper, we will show how to use our effective spin model
to study the zero temperature confinement in general and will implement such an
approach for the $(1+1)$ $SU(2)$ gauge field.

We shall first briefly introduce the effective spin model for the general $d+1$
dimensional finite temperature $SU(N)$ LGT. The finite temperature behavior of
the LGT can be studied by the following partition function on a hypercubic
lattice of size $N_{s}^{d}\times N_{\tau}$,
\begin{equation}
          Q = \sum e^{S}
\end{equation}
where $S$ is the Wilson action
\begin{equation}
       S = \beta_{E}\sum_{n} \sum_{\mu<\nu}
 \frac{1}{N} ReTr(U^{\mu}_{n}U^{\nu}_{n+\mu}
 U^{\mu^{\dagger}}_{n+\nu}
U^{\nu^{\dagger}}_{n})
\label{two}
\end{equation}
In eq.(\ref{two}), $\beta_{E}$ is the coupling constant, $n$, $\mu$ and
$\nu$ the space-time coordinate and directions respectively and $U$ the gauge
field link variables. Finite temperature can be introduced to the system by
imposing a periodic boundary condition with $N_{\tau}$ period in the time
direction. The temperature $T$ is then $\frac{1}{N_{\tau}a}$, where $a$ is the
lattice spacing. Naturally the zero temperature LGT can be obtained by simply
letting $T \rightarrow 0$.

To manifest the Wilson line variable in the action, we will introduce a thermal
gauge fixing for the gauge field.
\begin{eqnarray}
        U_{{\bf n},n_{\tau}}^{\tau} = 1 \hspace{0.2in}
(1 \leq n_{\tau} \leq N_{\tau}-1),
\end{eqnarray}
and the 
link $U_{{\bf n},N_{\tau}}^{\tau}$ remains unchanged. Noticed that we have
relabelled the space-time coordinate $n$ as ${\bf n}$ the space vector and
$n_{\tau}$ the time coordinate. With the thermal gauge, the trace of $U_{{\bf
n},n_{\tau}}^{\tau}$ becomes a Wilson line (labelled here as $W_{\bf n}$) and
is manifested in the action $S$. We shall rewrite the action $S$ as a sum of
$S^{g}$ and $S^{\tau}$.
\begin{equation}
        S = S^{g} + S^{\tau}
\end{equation}
where
\begin{eqnarray}
        S^{g} &=& \beta_{E}\sum_{{\bf n},i<j} \sum_{n_{\tau}=1}^{N_{\tau}}
 \frac{1}{N} ReTr(U^{i}_{{\bf n},n_{\tau}}U^{j}_{{\bf n}+i,n_{\tau}}
 U^{i^{\dagger}}_{{\bf n}+j,n_{\tau}}
U^{j^{\dagger}}_{{\bf n},n_{\tau}})\nonumber\\
          & & +\beta_{E} \sum_{{\bf n},i}\sum_{n_{\tau}=1}^{N_{\tau}-1}
 \frac{1}{N} ReTr(U^{i}_{{\bf n},n_{\tau}}
 U^{i^{\dagger}}_{{\bf n},n_{\tau}+1})
\label{four}
\end{eqnarray}
and
\begin{eqnarray}
       S^{\tau}   &=& \beta_{E} \sum_{{\bf n},i}
  \frac{1}{N} ReTr(U^{i}_{{\bf n},N_{\tau}}
    W_{{\bf n}+i}U^{i^{\dagger}}_{{\bf n},1}
  W^{\dagger}_{\bf n}).
\label{five}
\end{eqnarray}
Using eqs.(\ref{four}) and (\ref{five}), we can construct the effective action
in which the field variables are the space-like link and the Wilson line.

To derive the effective spin model, we will decouple the action into two parts:
one depends on the Wilson line and the other the space-like field:
\begin{equation}
       S_{eff} = S_{eff}^{W}(\{W_{\bf n}\}) +
                 S_{eff}^{U}(\{U^{i}_{{\bf n},n_{\tau}}\}),
\label{ff}
\end{equation}
In eq.(\ref{ff}), $S_{eff}^W$ and $S_{eff}^{U}$ are
\begin{eqnarray}
          S_{eff}^{W} &=& ln(\frac{\sum_{U_{n}^{i}}exp(S^{g}+S^{\tau})}
         {\sum_{U_{n}^{i}}exp(S_{eff}^{U})}) \\
          S_{eff}^{U} &=& ln(\frac{\sum_{W}exp(S^{g}+S^{\tau})}
         {\sum_{W}exp(S_{eff}^{W})})
\end{eqnarray}
In this way, we can keep the partition function of the action $S$ the same as
$S_{eff}$. Clearly, $S_{eff}^{U}$ and $S_{eff}^{W}$ will describe exactly the
behavior of $U$ and $W$ respectively.

Of course, $S_{eff}$ should preserve the symmetry of $S$ given by
eqs.(\ref{four}) and (\ref{five}). Since there is a local gauge invariance for
$W$ in $S$ (eqs.(\ref{four}) and (\ref{five})),
\begin{eqnarray}
         W_{\bf n}\rightarrow g_{\bf n}W_{\bf n}g_{\bf n}^{-1}, \hspace{0.2in}
   U^{i}_{{\bf n},n_{\tau}}\rightarrow g_{{\bf n},n_{\tau}}
        U^{i}_{{\bf n},n_{\tau}}g^{-1}_{{\bf n}+i,n_{\tau}}. \nonumber
\end{eqnarray}
where $g_{\bf n}$ and $g_{{\bf n},n_{\tau}}$ are the $SU(N)$ matrices,
$S_{eff}^{W}$ must depend only on $Tr(W^{m})$ (where m is an integer) and its
conjugate (from now on written as a function of $Tr(W^{m})$).
\begin{eqnarray}
       S_{eff}^{W}(\{W_{\bf n}\}) \equiv S_{eff}^{W}
        (\{\frac{1}{N}Tr(W_{\bf n}^{m})\})
\end{eqnarray}
We then use the variational principle to derive the exact form of $S^{W}_{eff}$
in Ref.\cite{zhu}
\begin{eqnarray}
     S_{eff}^{W} &=& \alpha \beta_{E} \sum_{{\bf n},i}
   Re (\frac{1}{N} TrW_{{\bf n}+i}\frac{1}{N}TrW^{\dagger}_{\bf n}),
\label{six}
\end{eqnarray}
where
\begin{equation}
     \alpha = Re(\langle\frac{1}{N}Tr(U^{i}_{{\bf n},N_{\tau}}
U^{i^{\dagger}}_{{\bf n},1})\rangle_{U}).
\label{seven}
\end{equation}
The symbol $\langle\cdot\cdot\cdot\rangle_{U}$ represents an average in
$S^{U}_{eff}$. To facilitate subsequent discussions, we shall redefine
\begin{eqnarray}
   \beta  \equiv \alpha \beta_{E}.
\label{eight}
\end{eqnarray}

To study $S_{eff}^{W}$ quantitively, $\beta$ must be computed. This means that
we need the exact form of $S^{U}_{eff}$. Physically, this also 
means that we should find 
$S^{U}_{eff}$ which can describe the $U$ link field behavior
exactly. To this end, we found that $S^{g}$ will describe the exact zero
temperature $U$ field behavior, and thus the important result that
$S^{U}_{eff}$ is simlpy $S^{g}$.
\begin{equation}
       S^{U}_{eff}(T=0) = S^{g}
\end{equation}
Once $S^{U}_{eff}$ is known, $\beta$ can be computed using eq.(\ref{seven}),
and we can now study the effective spin model.

\begin{figure}
\begin{picture}(400,170)(10,5)

\thicklines
\put(100,20){\vector(1,0){200}}
\put(100,20){\vector(0,1){150}}

\put(100,20){\rule{3cm}{1mm}}
\put(180,5){\bf $\beta_{c}$}
\put(55,100){\bf $\langle W \rangle_{S^{W}_{eff}}$}
\put(300,5){\bf $\beta$}

\end{picture}
\caption{The $\langle W \rangle_{S^{W}_{eff}}$ axis is the Wilson line. $\beta$
axis is the nearest-neighbor coupling of the effective spin model. The
thickline represents the $SU(N)$ confinement region and $\beta > \beta_{c}$ the
deconfinement region.}
\label{fig1}
\end{figure}
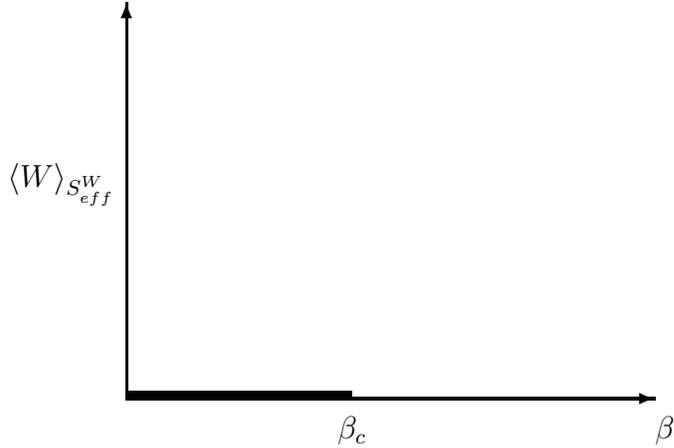

With the effective spin model in eq.(\ref{six}), the zero temperature $SU(N)$
confinement can now be studied. We shall take two steps to carry out such a
task. First, we shall calculate the nearest-neighbor coupling $\beta$ using
eqs.(\ref{seven}) and (\ref{eight}). Second, the $d$ dimensional effective spin
model of eq.(\ref{six}) will be examined numerically or analytically in order
to obtain the deconfinement-confinement phase diagram with respect to the
coupling constant $\beta$. To see the phase diagram clearly, let us define
the effective spin model critical coupling constant of the
deconfinement-confinement phase transition as $\beta_{c}$. We then plot the
phase diagram in Fig.\ref{fig1}. The comparison of the coupling constant
$\beta$ in the continuum limit with $\beta_{c}$ in Fig.\ref{fig1} can be used
as a criteria to determine whether the system confines or not.

Obviously, computing the continuum limit $\beta$ is the heart of the matter. If
it is possible, then we can determine whether the gauge field is confined or
not by "simply" placing $\beta$ in Fig.\ref{fig1}. For situations in which
$\beta$ cannot be computed exactly, two facts may still render the study
tractable. First it is clear that $\beta$ is a local space parameter for the
effective spin model (see eqs.(\ref{six}), (\ref{seven}) and (\ref{eight})).  Therefore,
it may be obtained perturbatively. Second, from Fig.\ref{fig1}, we see that
if the calculated upper limit of $\beta$ in the continuum limit
is less than $\beta_{c}$, then the non-Abelian gauge field confinement
is proved. These two points imply the feasibility of this approach.

We will now implement our idea on the $(1+1)$ dimensional $SU(2)$ gauge theory.
First we shall calculate $\beta$ by studying the $S^{U}_{eff}$. In the
fundamental representation, the parametrized link variable $U$ is

 \begin{equation}
        U = S^{0} + i \vec{\bf S} \cdot \vec{\sigma}
\end{equation}
where $S^{0}$ is a real number, $\vec{\bf S}$ a real vector and $\vec{\sigma}$
a Pauli matrix. Since $U$ is unitrary, $S^{0}$ and $\vec{\bf S}$ must satisfy
the following equation
\begin{equation}
         |S^{0}|^{2} + |\vec{\bf S}|^{2} = 1.
\end{equation}
We see that $S^{U}_{eff}$ is a one dimensional vector model with four internal
degrees of freedom, i.e.
\begin{equation}
      S^{U}_{eff}=\beta_{E}\sum_{i=1}^{N_{\tau}}\nS_{i}\cdot\nS_{i+1}
\end{equation}
where $\nS_{i}$ is a unit vector of four components, with $S^{0}_{i}$ and
$\vec{\bf S}_{i}$ as components. We find that this action can be exactly solved
\cite{solv} and from which one can obtain $\alpha$ as
\begin{equation}
         \alpha = (\frac{I_{2}(\beta_{E})}{I_{1}(\beta_{E})})^{N_{\tau}-1}
\end{equation}
where $I_{1}(\beta_{E})$ and $I_{2}(\beta_{E})$ are the modified Bessel
functions, and thus the nearest-neighbor coupling $\beta$ of the effective spin
model.
\begin{equation}
     \beta = \beta_{E}(\frac{I_{2}(\beta_{E})}{I_{1}(\beta_{E})})^{N_{\tau}-1}
\end{equation}
Since $I_{2}(\beta_{E})<I_{1}(\beta_{E})$ and $N_{\tau}$ approach infinity,
$\beta$ must be an infintesimally small number. For $\beta_{E}\rightarrow
\infty$ which is the continuum limit, we still have $\beta$ as
\begin{equation}
     \beta
 =\beta_{E}(\frac{(1-\frac{15}{8\beta_{E}}+O(\frac{1}{\beta_{E}^{2}}))}
           {(1-\frac{3}{8\beta_{E}}+O(\frac{1}{\beta_{E}^{2}}))})^{N_{\tau}-1}
\end{equation}
From the above one sees that $\beta$ is still infinitesimally small.
We therefore conclude
that $\beta$ is infinitesimally small for all $\beta_{E}$.

Once $\beta$ is known, we can then study the effective spin model. To this end,
let us parametrize $W$ as
\begin{equation}
           W = P^{0} + i \vec{\bf P}\cdot \vec{\sigma}
\end{equation}
where $P^{0}$ is real number and $\vec{\bf P}$ a real vector. Then the
$1$ dimensional effective spin model is given by
\begin{equation}
    S^{W}_{eff} = \beta \sum_{i} P^{0}_{i}P^{0}_{i+1}
\end{equation}
Obviously, for an infintesimally small $\beta$, $\langle Tr W
\rangle_{S^{W}_{eff}}$ vanishes. This is of course not surprising because the
$(1+1)$ dimensional $SU(2)$ gauge field is known to be permanently confined.

To show the consistency of our method with others, we shall use
this model to calculate the Polyakov static quark potential which is defined by
\begin{equation}
       V_{q\bar{q}}^{P}(R)= -\frac{1}{N_{\tau}} ln \frac{\langle \h TrW_{R} \h
 TrW_{0}^{\dagger}
\rangle_{S^{W}_{eff}}}{\langle \h TrW_{0} \h TrW_{0}^{\dagger}
 \rangle_{S^{W}_{eff}}}
\end{equation}
In the above, $R$ is the lattice distance between a quark $q$ and an antiquark
$\bar{q}$ and $\langle \cdots \rangle_{S^{W}_{eff}}$ represents an average in
the action $S^{W}_{eff}$. The $SU(2)$ group measure is
\begin{equation}
     [dW] = \frac{2}{\pi}\sqrt{1-(P^{0})^{2}} dP^{0}
\end{equation}
Since $\beta$ is infinitesimally small, this one dimensional effective spin
model can be solved by the strong coupling expansion. The results are
\begin{eqnarray}
       \langle \h TrW_{R} \h TrW_{0}^{\dagger}\rangle_{S^{W}_{eff}} &=&
                 \beta^{R} b^{R+1}(1+ O(\beta^{2})+ \cdots)
\end{eqnarray}
and
\begin{eqnarray}
       \langle \h TrW_{0} \h TrW_{0}^{\dagger} \rangle_{S^{W}_{eff}} &=&
         b (1+ O(\beta^{2}) + \cdots )
\end{eqnarray}
where $b$ is
\begin{equation}
      b = \int_{-1}^{1} dP^{0} (P^{0})^{2} \frac{2}{\pi} \sqrt{1-(P^{0})^{2}}
\end{equation}
From this one can readily show that the Polyakov static quark antiquark
potential is a linear one
\begin{equation}
       V_{q\bar{q}}^{P}(R) = Rln(\frac{I_{1}(\beta_{E})}{I_{2}(\beta_{E})})
\end{equation}
Therefore we have shown that the confinement (here we mean that the Wilson
line vanishes.) corresponds to the linear Polyakov quark antiquark potential.

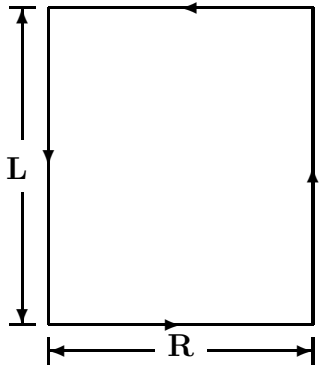
\begin{figure}

\begin{picture}(400,150)(10,5)

\thicklines
\put(150,20){\line(1,0){100}}
\put(150,20){\vector(1,0){50}}
\put(150,140){\line(1,0){100}}
\put(250,140){\vector(-1,0){50}}
\put(150,20){\line(0,1){120}}
\put(150,140){\vector(0,-1){60}}
\put(250,20){\line(0,1){120}}
\put(250,20){\vector(0,1){60}}

\put(150,15){\line(0,-1){10}}
\put(250,15){\line(0,-1){10}}
\put(190,10){\vector(-1,0){40}}
\put(210,10){\vector(1,0){40}}
\put(195,8){\bf R}

\put(145,20){\line(-1,0){10}}
\put(145,140){\line(-1,0){10}}
\put(140,70){\vector(0,-1){50}}
\put(140,90){\vector(0,1){50}}
\put(134,75){\bf L}

\end{picture}
\caption{The $R \times L$ Wilson loop.}
\label{fig2}
\end{figure}

Next we will also calculate the Wilson potential by studying the $R\times L$
Wilson loop in Fig.\ref{fig2}. Here $L$ and $R$ are the lattice distance in the
time and space directions respectively. The quark antiquark potential in the
action $S^{g}$ can be defined as
\begin{equation}
           V_{q\bar{q}}^{W}(R) = -\lim_{L \rightarrow \infty}\frac{1}{L}
         ln \langle W_{RL} \rangle_{S^{g}}
\end{equation}
Here $W_{RL}$ is the Wilson loop, $ V_{q\bar{q}}^{W}(R)$ is referred to as the
Wilson potential and $\langle \cdots \rangle_{S^{g}}$ represents an average
in the action $S^{g}$. We can then simply obtain $W_{RL}$ in the  $S^{g}$ as
\begin{equation}
     \langle W_{RL} \rangle_{S^{g}} =
 (\frac{I_{2}(\beta_{E})}{I_{1}(\beta_{E})})^{RL},
\end{equation}
from which the Wilson potential
\begin{equation}
      V_{q\bar{q}}^{W}(R) = Rln(\frac{I_{1}(\beta_{E})}{I_{2}(\beta_{E})})
\end{equation}
is obtained. Clearly, the zero temperature Polyakov potential
$V_{q\bar{q}}^{P}(R)$ is identical to the Wilson potential
$V_{q\bar{q}}^{W}(R)$.

In summary, a new vista to study the zero temperature $SU(N)$ confinement has
been introduced. We have solved the effective spin model of the $(1+1)$ $SU(2)$
gauge field exactly, and from which its confinement and linear Polyakov quark
antiquark potential are shown. Also for the first time the Polyakov potential
is shown to be identical to the Wilson potential. For $(3+1)$ $SU(3)$ gauge
field, the study will certainly be more complicated. However, since our
analysis have shown that the effective spin model coupling constant is a local
space parameter and thus can be computed perturbatively, it is hoped that
this approach can lead to a resolution of the important question of the $(3+1)$
$SU(3)$ confinement in the continuum limit. 

\begin{flushleft}
$^{\star}$ This work is supported by the National Science Foundation.
\end{flushleft}


\begin{thebibliography}{99}
\bibitem[1]{creutz} For example,
M. Creutz, Phys. Rev., {\bf D21}, 2308(1980), \\
                    M. Creutz, Phys. Rev. Lett., {\bf 45}, 313(1980).
\bibitem[2]{seiler}E. Seiler, in {\em Quark Confinement and Liberation},
edited by F. R. Klinkhamer and M. B. Halpern (World Scientific, Singapore,
1985).
\bibitem[3]{ex1}The Wilson line has been used to study the Polyakov
quark antiquark potential. See for example,
 M. Teper, in {\em Quark Confinement and Liberation},
edited by F. R. Klinkhamer and M. B. Halpern (World Scientific, Singapore,
1985).
\bibitem[4]{zhu} Z. Zhu and D. H. Feng, Modern Physics Letter, {\bf A}, Submitted.\\
   Z. Zhu and D. H. Feng, Phys. Rev., {\bf D48}, 397(1993).          
\bibitem[5]{solv}H. E. Stanley, Phys. Rev. {\bf 179}, 570(1969).
\end{thebibliography}
\end{document}